\documentclass[%
superscriptaddress,
reprint,
preprintnumbers,
amsmath,amssymb,aps,prl]{revtex4-1}

\usepackage{bm}
\usepackage[colorlinks=true,linktocpage=true,linkcolor=blue,citecolor=blue,urlcolor=blue]{hyperref}

\usepackage{cleveref} 

\usepackage[utf8]{inputenc} 
\usepackage[T1]{fontenc} 
\usepackage{microtype}

\usepackage[mathscr]{euscript}

\usepackage{tensor}  

\usepackage{color,soul}

\newcommand{\half}{\frac{1}{2}}
\newcommand{\nn}{\nonumber}

\newcommand{\lie}{\pounds}

\newcommand{\dow}{\partial}

\usepackage{xparse}

\ExplSyntaxOn

\clist_map_inline:nn{A,B,C,D,E,F,G,H,I,J,K,L,M,N,O,P,Q,R,S,T,U,V,W,X,Y,Z}
{
  \expandafter\def\csname b#1\endcsname{\bm{#1}} 
  
  \expandafter\def\csname c#1\endcsname{\mathcal{#1}} 
  
  \expandafter\def\csname bc#1\endcsname{\bm{\mathcal{#1}}} 
  
  \expandafter\def\csname s#1\endcsname{{\mathsmaller{#1}}} 
  
  \expandafter\def\csname bb#1\endcsname{\mathbb{#1}} 
  
  \expandafter\def\csname rm#1\endcsname{\mathrm{#1}} 
  
  \expandafter\def\csname sc#1\endcsname{\mathscr{#1}} 
  
  \expandafter\def\csname sf#1\endcsname{\mathsf{#1}} 
  
  \expandafter\def\csname f#1\endcsname{\mathfrak{#1}} 
}

\ExplSyntaxOff

\newcommand{\lb}{\left (}
\newcommand{\rb}{\right )}

\newcommand\ext{\text{ext}}

\begin{document} 

\preprint{DCPT-18/25}

\title{Magnetohydrodynamics as superfluidity}

\author{Jay Armas} 
\email{j.armas@uva.nl}
\affiliation{Institute for Theoretical Physics, University of Amsterdam,
  1090 GL Amsterdam, The Netherlands}
\affiliation{Dutch Institute for Emergent Phenomena, The Netherlands}

\author{Akash Jain}
\email{akash.jain@durham.ac.uk} 
\affiliation{Centre for Particle Theory \& Dept.\ of Mathematical Sciences,
  Durham University, Durham DH1 3LE, United Kingdom} 


\begin{abstract}
  We show that relativistic magnetohydrodynamics (MHD) can be recast as a novel
  theory of superfluidity. This new theory formulates MHD just in terms of
  conservation equations, including dissipative effects, by introducing
  appropriate variables such as a magnetic scalar potential, and providing
  necessary and sufficient conditions to obtain equilibrium configurations. We
  show that this scalar potential can be interpreted as a Goldstone mode
  originating from the spontaneous breaking of a one-form symmetry, and present
  the most generic constitutive relations at one derivative order for a
  parity-preserving plasma in this new superfluid formulation.
\end{abstract}

\pacs{Valid PACS appear here}

\maketitle


Relativistic magnetohydrodynamics (MHD) provides a universal framework to study
plasma physics in astrophysical settings as well as in laboratory
experiments~\cite{goedbloed2004principles}.  As an effective theory for
interactions between electromagnetic and thermal degrees of freedom of matter,
MHD describes the coupling of Maxwell's equations to hydrodynamics.  For most of
its applications, MHD is formulated under the assumption that the electric
fields are Debye-screened and are therefore weak/short-ranged, while the
magnetic fields could be arbitrary, and that the plasma is electrically neutral
at hydrodynamic length scales. Formulations of MHD under these assumptions have
been extensively studied and have a wide range of applicability. However, the
structural foundations and transport properties of MHD have only recently
received considerable attention~\cite{Schubring:2014iwa, Grozdanov:2016tdf,
  Hernandez:2017mch, Armas:2018ibg}. The main purpose of this letter is to
present an equivalent formulation of dissipative MHD as a system of
(higher-form) conservation equations, which is better suited for numerical
analyses, and to resolve certain underlying technical issues in the definition
of hydrostatic equilibrium.

Traditional treatments of MHD are formulated in terms of a stress tensor
$T^{\mu\nu}$ and charge current $J^{\mu}$, subject to energy-momentum
conservation, Maxwell's equations, and Bianchi identity
(e.g. see~\cite{Hernandez:2017mch})
\begin{equation}\label{EOM_MHD}
\begin{gathered}
  \nabla_\mu T^{\mu\nu} = F^{\nu\rho} J_\rho,~
  J^\mu + J^{\mu}_{\text{ext}} = 0,~
  \epsilon^{\mu\nu\rho\sigma}\nabla_\nu F_{\rho\sigma}=0~.
\end{gathered}
\end{equation}
Here the components of $F_{\mu\nu}$ are the electromagnetic fields and
$J^\mu_\ext$ is a conserved external charge source (e.g. a lattice of ions)
satisfying $\nabla_\mu J^\mu_\ext = 0$. The charge current $J^\mu$ can be split
into $\nabla_\nu F^{\nu\mu} + J^\mu_{\text{matter}}$, where
$J^\mu_{\text{matter}}$ is the contribution from matter fields, converting
Maxwell's equations into their better known form. The Bianchi identity is solved
by introducing the photon field such that
$F_{\mu\nu} = 2\partial_{[\mu}A_{\nu]}$, while the remaining 8 equations govern
the dynamics of the temperature $T$, chemical potential $\mu$, fluid velocity
$u^\mu$, and gauge field $A_\mu$ \cite{Note1}. In usual MHD applications,
$J^\mu _\protect \text {ext}$ is taken to be zero while $\mu $ is fixed such
that the electric charge vanishes at hydrodynamic length scales.
 
In practise, however, for instance in numerical approaches, Maxwell's
equations can be used to eliminate $\mu$ and Debye-screened electric fields
$E_\mu = F_{\mu\nu} u^\nu$ from the system, leaving $u^\mu$, $T$, and magnetic
fields $B^\mu = \half \epsilon^{\mu\nu\rho\sigma} u_\nu F_{\rho\sigma}$ to be
the only relevant fields. The dynamics for $u^\mu$ and $T$ is governed by 
energy-momentum conservation, while that for $B^\mu$ by the Bianchi
identity (see e.g.~\cite{Gammie:2003rj,Font2008}).

Extrapolating this line of thought, the authors of~\cite{Grozdanov:2016tdf} (see
also~\cite{Schubring:2014iwa}), inspired by the framework of generalised global
symmetries~\cite{Gaiotto:2014kfa}, proposed a formulation of MHD in terms of
string/one-form fluids \cite{Caldarelli:2010xz, Schubring:2014iwa,
  Gaiotto:2014kfa, Grozdanov:2016tdf, Hernandez:2017mch, Armas:2018ibg}. The key
observation is that once Maxwell's equations in~\eqref{EOM_MHD} are
explicitly solved by setting $J^\mu = - J^\mu_\ext$, the dynamics of MHD
effectively reduces to that of a fluid with a global one-form symmetry. To wit,
by defining a two-form current
$J^{\mu\nu} = \half \epsilon^{\mu\nu\rho\sigma}F_{\rho\sigma}$, and identifying
the external charge source as
$J^\mu_\ext = \frac16 \epsilon^{\mu\nu\rho\sigma}H_{\nu\rho\sigma}$, where
$H_{\mu\nu\rho} = 3 \dow_{[\mu}b_{\nu\rho]}$ is the field strength associated
with a background two-form gauge field $b_{\mu\nu}$, \cref{EOM_MHD} can be
rewritten as
\begin{equation}\label{EOM_dual}
  \nabla_{\mu} T^{\mu\nu} = \half H^{\nu\rho\sigma} J_{\rho\sigma}~,~~
  \nabla_{\mu} J^{\mu\nu} = 0~.
\end{equation}
The dynamical variables of MHD in this string fluid formulation are the string
chemical potential $\varpi$ and the vector characterising the direction of
strings $h^\mu$ (with $h^\mu h_\mu = 1$ and $u^\mu h_\mu = 0$), in addition to
$u^\mu$ and $T$. Heuristically, $h^\mu$ corresponds to the direction of magnetic
fields, while $\varpi$ is a chemical potential conjugate to their strength
\begin{equation}
  \varpi = - 2 |B| \frac{\dow P}{\dow B^2} + \mathcal{O}(\dow)~,~~
  h^\mu = \frac{B^\mu}{|B|} + \mathcal{O}(\dow)~,
\end{equation}
where $P(T,B^2)$ is the pressure of MHD. The string fluid pressure defined later
in \cref{GHI_ideal_consti} is related to MHD pressure as
$p(T,\varpi) = P(T,B^2) - 2|B|^2 \dow P/\dow B^2 + \mathcal{O}(\dow)$. These
relations admit corrections at higher derivative orders.

If we switch off the sources $H_{\mu\nu\lambda}=0$, \cref{EOM_dual} with the
ideal order constitutive relations \eqref{GHI_ideal_consti} and equation of
state $P(T,B^2) = P(T) - \half B^2$ or $p(T,\varpi) = P(T) + \half \varpi^2$
reduces to the system of equations given in~\cite{Gammie:2003rj}. The first
equation is the well-known energy-momentum conservation. The spatial components
of the second equation can be seen as the induction equation, while the
time-component as the no-monopole constraint of~\cite{Gammie:2003rj}.

As pointed out in~\cite{Armas:2018ibg}, the string fluid variables $\varpi$ and
$h^\mu$, while consistent, are not well suited for describing equilibrium
configurations in MHD. In particular, a generic string fluid equilibrium
configuration cannot be derived from a hydrostatic partition function within the
framework of~\cite{Grozdanov:2016tdf}.  Such equilibrium configurations serve as
initial conditions in numerical simulations of hydrodynamics, so it is crucial
that we identify the appropriately suited degrees of freedom. In this letter, we
introduce a more natural pair of fields: gauge-non-invariant one-form chemical
potential $\mu_\mu$ and ``scalar Goldstone'' $\varphi$ such that
\begin{equation}\label{eq:hm}
  \varpi h_\mu = \mu_\mu - T\dow_\mu \varphi.
\end{equation}
When coupled to a time-independent background, we find that in equilibrium
$\mu_\mu/T = b_{t\mu}$, while $\varphi$ plays a role similar to that of a
``magnetic scalar potential'' and is solved for using the no-monopole
constraint. Drawing a comparison with the Goldstone phase field in typical
superfluids, we formulate a novel theory of \emph{one-form superfluidity}, where
the underlying global one-form symmetry is spontaneously broken leading to a
one-form Goldstone mode $\varphi_\mu$~\cite{Gaiotto:2014kfa, Lake:2018dqm,
  Hofman:2018lfz}. We show that the existence of this mode gives rise to a
well-defined hydrostatic sector for string fluids, when viewed as a limit of
one-form superfluids where only a part of the one-form symmetry is broken, with
the associated scalar Goldstone $\varphi = u^\mu \varphi_\mu/T$.

Finally, while the traditional and string fluid formulations of MHD can easily
be shown to be equivalent at ideal order as described above, at higher
derivative orders this equivalence is quite non-trivial. It has only been
established in the dissipative sector for linear fluctuations (Kubo formulae) in
a state with $\mu = 0$~\cite{Hernandez:2017mch}. We show an exact correspondence
between MHD and our improved formulation of string fluids. The crucial
ingredients of this correspondence are presented in this letter, while further
details are relegated to a companion publication~\cite{Armas:2018zbe}.

\vspace{1em} \emph{String fluids and equilibrium.}---One-form hydrodynamics is
governed by the equations of motion \eqref{EOM_dual} and respective constitutive
relations, that is, the most generic expressions for $(T^{\mu\nu}, J^{\mu\nu})$
in terms of $(T,u^\mu,\varpi,h^\mu)$ and $(g_{\mu\nu},b_{\mu\nu})$ allowed by
symmetries and the second law of thermodynamics. At ideal order, these relations
read
\begin{align}\label{GHI_ideal_consti}
  T^{\mu\nu}
  &= (\epsilon+p) u^\mu u^\nu + p\, g^{\mu\nu} - \varpi \rho\, h^\mu h^\nu
    + \mathcal{O}(\dow)~, \nn\\
  J^{\mu\nu}
  &= 2 \rho\, u^{[\mu}h^{\nu]} + \mathcal{O}(\dow)~,
\end{align}
where $p(T,\varpi)$ is an arbitrary function, while $\epsilon(T,\varpi)$ and
$\rho(T,\varpi)$ are determined by the thermodynamic relations
$\epsilon + p = T s + \varpi \rho$ and $dp = sdT + \rho d \varpi$.  The
associated entropy current $S^\mu = s u^\mu$ is trivially conserved (see
\cite{Grozdanov:2016tdf} for more details).

Hydrodynamics is the study of small fluctuations of a quantum system around
thermodynamic equilibrium and hence it is important to understand how to
describe equilibrium configurations.  As usual, one assumes the existence of an
arbitrary time coordinate $t$ such that
$g_{tt} < 0,~\dow_t g_{\mu\nu} = \dow_t b_{\mu\nu} = 0$, and
$u^{\mu}/T = \delta^\mu_t$.  However, this is not sufficient to attain
equilibrium in string fluids since the conservation of string charge
$\nabla_\mu\left(T\rho h^\mu\right)=0$, arising from the no-monopole constraint
in \cref{EOM_dual}, is not satisfied.  To circumvent this issue, the authors
of~\cite{Grozdanov:2016tdf} specialised to backgrounds that further admit a
spatial coordinate $z$ such that $\dow_z g_{\mu\nu} = \dow_z b_{\mu\nu} = 0$
with $g_{tz} = 0$ and obtained an equilibrium solution by setting
$h^{\mu}=\delta^\mu_z/\sqrt{g_{zz}}$ and $\varpi/T =
b_{tz}/\sqrt{g_{zz}}$. However, this ``equilibrium solution'', besides being
only a subset of the solutions to $\nabla_\mu\left(T\rho h^\mu\right)=0$,
generically contributes to entropy production~\cite{Armas:2018ibg}, which an
equilibrium fluid configuration, by definition, cannot. For example, consider a
particular dissipative correction to $J^{\mu\nu}$ obtained in
\cite{Grozdanov:2016tdf}
\begin{equation}\label{the_problematic_term}
  \delta J^{\mu\nu}_{(1)} \ni
  - r_{||} \Delta^{\mu\rho} \Delta^{\nu\sigma}
  \left[ 2T \nabla_{[\rho}\left(\frac{\varpi}{T} h_{\sigma]} \right)
    + u_\lambda {H^\lambda}_{\rho\sigma} \right]~.
\end{equation}
Here $r_{||}\geq0$ is a dissipative transport coefficient and
$\Delta^{\mu\nu}=g^{\mu\nu}+u^{\mu}u^{\nu}-h^{\mu}h^{\nu}$. It may be explicitly
checked that this term does not vanish when evaluated on the equilibrium
solution of~\cite{Grozdanov:2016tdf}. Therefore, it must be imposed to vanish by
hand as an ad-hoc constraint on equilibrium backgrounds, in addition to
requiring an isometry along the coordinate $z$. An infinite cascade of similar
conditions show up at every derivative order~\cite{Armas:2018ibg}.
Thus, in contrast to the case of typical charged fluids, the hydrostatic sector
of string fluids is ill-defined. We show that there is a first-principle
derivation of equilibrium configurations that do not require 
ad-hoc
constraints nor 
 existence of a preferred coordinate $z$.

\vspace{1em} \emph{Revisiting string fluids.}---Some of the issues mentioned
above have an 
analogue in superfluid dynamics. In this context, had we
considered the components of the superfluid velocity $\xi^\mu$ as fundamental
degrees of freedom, ignorant of its definition
$\xi_\mu= \partial_\mu\phi + A_\mu$ in terms of the Goldstone mode $\phi$, we
could be tempted to introduce a preferred $z$-coordinate in equilibrium to
align $\xi^\mu$ with. However, it is precisely $\phi$ that leads to
well-defined equilibrium configurations for
superfluids~\cite{Bhattacharyya:2012xi}.

These considerations lead us to re-evaluate whether the string fluid variables
$(T,u^{\mu},\varpi, h^\mu)$ describe a symmetry-unbroken phase or if the
underlying one-form symmetry is spontaneously broken. In order to identify the
correct hydrodynamic fields in a symmetry-unbroken phase, we follow the approach
of~\cite{Haehl:2015pja} for usual charged fluids.  In this setting, the fields
$(T,u^\mu,\mu)$ can be exchanged by a set of symmetry parameters
$\mathscr{B} = (\beta^\mu, \Lambda^\beta)$, where $\beta^\mu=u^{\mu}/T$ and
$\Lambda^\beta=\mu/T-\beta^\mu A_\mu$.  Under the action of an infinitesimal
symmetry transformation $\mathscr{X} = (\chi^\mu,\Lambda^\chi)$, with $\chi^\mu$
being a diffeomorphism and $\Lambda^\chi$ a gauge transformation, they transform
according to $\delta_\scX \beta^\mu = \lie_\chi \beta^\mu$ and
$\delta_\scX \Lambda^\beta = \lie_\chi \Lambda^\beta - \lie_\beta
\Lambda^\chi$. It may be explicitly checked that $\delta_\scX \mu=\lie_\chi \mu$
and therefore that $\mu$ is gauge-invariant. Motivated by this, in the
symmetry-unbroken phase of string fluids we consider the fields
$\mathscr{B} = (\beta^\mu, \Lambda^\beta_\mu)$ and introduce the one-form
chemical potential $\mu_\mu$ via the relation
\begin{equation} \label{eq_mu}
  \frac{\mu_\mu}{T} = \Lambda^\beta_\mu + \beta^\nu b_{\nu\mu}~~.
\end{equation}
Given the transformation property $ \delta_{\scX} \Lambda^{\beta}_{\mu}=
\lie_{\chi} \Lambda^{\beta}_{\mu} - \lie_{\beta}
\Lambda^{\chi}_{\mu}$ under the action of $\scX = (\chi^\mu,
\Lambda^\chi_\mu)$, it is straightforward to check that $\delta_\scX \mu_\mu =
\lie_\chi \mu_\mu - T\dow_\mu\lb \beta^\nu \Lambda^\chi_\nu
\rb$. Hence, unlike usual charged fluids,
$\mu_\mu$ is not gauge invariant and cannot correspond to $(\varpi,
h^\mu)$ of string fluids.
Specifically, we cannot construct a gauge-invariant vector that would replace
$h^\mu$ in~\cref{GHI_ideal_consti} using just $\mu_\mu$.

In order to construct a gauge-invariant vector, we need to introduce a scalar
field $\varphi$ that transforms in the non-trivial manner
$\delta_\scX \varphi = \chi^\mu\dow_\mu \varphi - \beta^\mu
\Lambda^\chi_\mu$. This allows for the definition of the gauge-invariant
combination given in \cref{eq:hm}.
The scalar $\varphi$ is accompanied by its own equation of motion, which,
following \cite{Jain:2016rlz}, reads
$\delta_{\scB}\varphi=\mathcal{O}\left(\dow\right)$ implying that
$u^\mu h_\mu=\mathcal{O}\left(\dow\right)$. Using a part of the redefinition
freedom in $\mu_\mu$, one may set $u^\mu h_\mu=0$ exactly, thus reproducing the
variables of string fluids. We learn that instead of treating $\varpi$ and
$h^\mu$ as fundamental hydrodynamic variables in string fluids, we should
instead work with $\mu_\mu$ and $\varphi$. This also leads to well-defined
equilibrium configurations given by $u^\mu/T = \delta^\mu_t$ and
$\mu_\mu/T = b_{t\mu}$ along with $\varphi = \varphi_0$ which solves the
expected ideal order Poisson's equation $\nabla_\mu\left(T\rho h^\mu\right)=0$,
i.e. the no-monopole constraint of \cite{Gammie:2003rj}
\begin{equation}\label{vf-EOM}
  \frac{1}{\sqrt{-g}} \dow_\mu \lb \sqrt{-g} \frac{\rho T^2}{\varpi}
  g^{\mu\nu} \lb b_{t\nu} - \dow_\nu \varphi_0 \rb \rb = \mathcal{O}(\dow^2)~.
\end{equation}
On this solution, the term in \cref{the_problematic_term} vanishes. The
introduction of $\varphi$ through \eqref{eq:hm} provides an improved version of
the string fluids formulated in \cite{Grozdanov:2016tdf}.

The transformation of $\varphi$ under the action of $\scX$ involves
the hydrodynamic field $\beta^\mu$, suggesting that, unlike usual superfluids,
$\varphi$ is not a fundamental variable. In fact, in equilibrium, under a gauge
transformation the scalar field transforms as
$\varphi\to\varphi-\Lambda_t^\chi$, hinting that $\varphi$ might be better
understood as the time component of a vector Goldstone mode $\varphi_\mu$,
embedded into a larger theory in which the one-form symmetry is spontaneously
broken.

\vspace{1em} \emph{One-form superfluids.}---As pointed out in
\cite{Gaiotto:2014kfa, Lake:2018dqm, Hofman:2018lfz}, the Goldstone mode
corresponding to the spontaneous breaking of a one-form symmetry is a dynamical
$\rmU(1)$ gauge field $\varphi_\mu$. Under the action of the set of one-form
symmetry parameters $\scX$, the Goldstone $\varphi_\mu$ transforms analogous to
its zero-form counterpart
\begin{equation} \label{eq:tp}
\delta_\scX \varphi_\mu= \lie_{\chi} \varphi_\mu-\Lambda^\chi_\mu~~.
\end{equation}
Thus, the scalar $\varphi$ appearing in \eqref{eq:hm} is in fact given by
$\varphi = \beta^\mu \varphi_\mu$.  We can define the gauge-invariant covariant
derivative of $\varphi_\mu$
\begin{equation}
  \xi_{\mu\nu}=2\partial_{[\mu}\varphi_{\nu]}+b_{\mu\nu}~,
\end{equation}
which is a higher-form analogue of the superfluid velocity and transforms simply
as $\delta_\scX \xi_{\mu\nu}= \lie_{\chi}\xi_{\mu\nu}$.

The dynamics of one-form superfluids is also governed by \cref{EOM_dual} with
constitutive relations written in terms of the hydrodynamic fields
$T,u^\mu,\xi_{\mu\nu}$, and $\varpi h_\mu$ defined in \cref{eq:hm}, supplemented
with the equation of motion for $\varphi_\mu$.  Formulating the offshell second
law of thermodynamics analogous to zero-form superfluids~\cite{Jain:2016rlz},
one can straightforwardly derive this equation at ideal order
\begin{equation} \label{eq_jose}
  u^\mu\xi_{\mu\nu} = \varpi h_\nu + \mathcal{O}\left(\dow\right)~~.
\end{equation}
This is a higher-form analogue of the Josephson equation for superfluids. Note
that the condition $u^\mu h_\mu = \mathcal{O}(\dow)$ of string fluids follows
from here. Using \cref{eq_jose},
we can remove $\varpi h_\mu$ from the independent set of
hydrodynamic variables in favour of $\zeta_\mu = \xi_{\mu\nu}u^\nu$. Hence, the
dynamics of one-form superfluids is governed by \eqref{EOM_dual} alone, along
with the offshell second law of thermodynamics
\begin{equation} \label{eq_fn}
  \nabla_\mu N^\mu=\frac{1}{2}T^{\mu\nu}\delta_{\scB}g_{\mu\nu}
  +\frac{1}{2}J^{\mu\nu}\delta_{\scB}\xi_{\mu\nu}+\Delta~,\quad
  \Delta\ge0~~.
\end{equation}
Here
$N^\mu = S^\mu + \frac{1}{T} T^{\mu\nu}u_\nu - \frac{1}{T} J^{\mu\nu}\zeta_\nu$
is the free energy current and
\begin{equation}
  \delta_{\scB}g_{\mu\nu} = 2\nabla_{(\mu}\beta_{\nu)},~
  \delta_{\scB}\xi_{\mu\nu} =
  - 2\nabla_{[\mu}\left(\zeta_{\nu]}/T \right)
  + \beta^\rho H_{\rho\mu\nu}~.
\end{equation}
\Cref{eq_fn} requires that for a given set of constitutive relations
$(T^{\mu\nu}, J^{\mu\nu})$ in terms of $(T,u^\mu,\xi_{\mu\nu},g_{\mu\nu})$,
there must exist a free energy current $N^\mu$ and a positive semi-definite
quadratic form $\Delta$ such that \cref{eq_fn} is satisfied.

In order to discuss the constitutive relations of a one-form superfluid in four
spacetime dimensions, we decompose
$\xi_{\mu\nu} = 2u_{[\mu}\zeta_{\nu]} -
\epsilon_{\mu\nu\rho\sigma}u^{\rho}\bar\zeta^{\sigma}$. A generic one-form
superfluid can depend on both $\zeta_\mu$ and $\bar\zeta_\mu$ arbitrarily but
here we mention two special cases that find a direct application in plasma
physics. The ``string fluid limit'' is the case in which the constitutive
relations depend on $\zeta_\mu = -\varpi h_\mu$ but not on $\bar\zeta_\mu$,
which, as we show below, is dual to MHD. Formally, removing $\bar\zeta_\mu$ from
the constitutive relations means that the one-form symmetry is only broken along
the timelike direction $\beta^\mu$ while the spatial part of the symmetry is
left intact. This gives rise to the improved string fluid theory described
earlier where $\varphi=\beta^\mu\varphi_\mu$ is introduced according to
\cref{eq:hm}.
Another interesting case
is the ``electric limit'', in which the hierarchy of gradients
$\zeta_\mu=\mathcal{O}\left(1\right)$ and
$\bar\zeta_\mu=\mathcal{O}\left(\dow\right)$ is assumed. This latter case, where
the full one-form symmetry is broken, can be shown to be equivalent to plasma in
the absence of free charges \cite{Armas:2018zbe}. Focusing on the former string
fluid limit, we note that there are two Lorentz and gauge invariant scalars at
ideal order, namely, $T$ and $\varpi = \sqrt{\zeta_\mu\zeta^\mu}$, on which the
free energy current $N^\mu=N(T,\varpi)\beta^\mu$ can depend. Using \eqref{eq_fn}
we find the ideal order constitutive relations~\eqref{GHI_ideal_consti}.  Thus,
at ideal order, one-form superfluids in this limit reduce to string fluid
dynamics, which continues to be the case at higher-orders.

\vspace{1em} \emph{One-derivative string fluids.}---
We parametrise the
non-hydrostatic corrections to the one-form superfluid constitutive relations as
\begin{align}\label{nhs-ansatz}
  \delta T^{\mu\nu}_{(1)}
  &= \delta f \Delta^{\mu\nu}+\delta \tau
    h^{\mu}h^{\nu}+2\ell^{(\mu}h^{\nu)}+t^{\mu\nu}~~, \nn\\
  \delta J^{\mu\nu}_{(1)}
  &= 2 m^{[\mu} h^{\nu]} + s^{\mu\nu}.
\end{align}
All the tensor structures appearing here are transverse to $u^\mu$ and $h^\mu$.
Recall that we had used 
part of the redefinition freedom in $\mu_\mu$
around \cref{eq:hm} to set $u^\mu h_\mu = 0$. In writing \cref{nhs-ansatz}, we
also used the residual freedom in $\mu_\mu$ along with that in $u^\mu$
and $T$ to work in an analogue of the ``Landau frame'' and set
$u_\mu \delta T^{\mu\nu}_{(1)} = u_\mu \delta J^{\mu\nu}_{(1)} = 0$. When
working with full one-form superfluids, we can instead choose to use the
redefinition of $\mu_\mu$ to make \cref{eq_jose} exact, at the expense of having
$u_\mu \delta J^{\mu\nu}_{(1)} \neq 0$~\cite{Armas:2018zbe}.

Restring ourselves to derivative corrections that respect CP invariance
\cite{Note2}, in the
non-hydrostatic sector we find
\begin{align}\label{eq:transport}
  \delta f
  &= - T/2 \lb \zeta_\perp \Delta^{\mu\nu}
  + \zeta_\times h^\mu h^\nu \rb \delta_{\scB}g_{\mu\nu}, \nn\\
  \delta \tau
  &= - T/2
  \lb \zeta'_\times \Delta^{\mu\nu}
  + \zeta_\parallel h^\mu h^\nu \rb \delta_{\scB}g_{\mu\nu}~, \nn\\
  \ell^\mu
  &= - T
  \lb \eta_\parallel \Delta^{\mu\sigma} + \tilde\eta_\parallel \epsilon^{\mu\sigma}\rb
    h^{\nu}\delta_{\scB}g_{\sigma\nu}~, \nn\\
  m^\mu
  &= - T
  \lb r_\perp \Delta^{\mu\sigma} + \tilde r_\perp \epsilon^{\mu\sigma}\rb
    h^{\nu}\delta_{\scB}\xi_{\sigma\nu}~, \nn\\
  t^{\mu\nu}
  &= - T \lb \eta_{\perp} \Delta^{\rho\langle\mu}
  \delta_{\scB} g_{\rho\sigma}
  - \tilde\eta_{\perp} \epsilon^{\rho\langle\mu} \rb \Delta^{\nu\rangle\sigma} 
    \delta_{\scB} g_{\rho\sigma}~, \nn\\
  s^{\mu\nu}
  &= - T r_{||}\Delta^{\mu\lambda}\Delta^{\nu\sigma}
    \delta_{\scB}\xi_{\lambda\sigma}~,
\end{align}
where $\epsilon^{\mu\nu} = \epsilon^{\mu\nu\rho\sigma}u_\rho h_\sigma$.  Using
\eqref{eq_fn}, we obtain exactly the same constraints
and number of dissipative transport coefficients as for string fluids 
I 
in
\cite{Grozdanov:2016tdf, Hernandez:2017mch}.

The hydrostatic sector of the theory
has not been considered
in~\cite{Grozdanov:2016tdf, Hernandez:2017mch}. This sector is described by a
hydrostatic effective action for the Goldstone mode.  Aligning the fluid velocity with
a timelike Killing vector, up to first order in derivatives, this action is
\begin{equation} \label{eq_part}
  \begin{split}
    \mathcal{S} = \int d^4x \sqrt{{-}g}
    \Big[ p
    - \frac{\alpha}{6}\epsilon^{\mu\nu\lambda\sigma}u_\mu H_{\nu\lambda\sigma}
    - \beta \epsilon^{\mu\nu} \partial_\mu u_\nu \Big],
  \end{split}
\end{equation}
where $\alpha(T,\varpi)$ and $\beta(T,\varpi)$ are hydrostatic transport
coefficients. Contrary to \cite{Grozdanov:2016tdf, Armas:2018ibg}, no
assumptions regarding the presence of 
spatial isometries in the
background are necessary. Extremising \cref{eq_part} with respect to $\varphi$
yields its equation of motion in equilibrium, which 
improves
\cref{vf-EOM} due to the 
$\alpha$ and
$\beta$. Eq.~\eqref{eq_part} characterises all equilibrium configurations of MHD
and guarantees that all non-hydrostatic contributions in \cref{eq:transport}
vanish.  We will now show that one-form superfluidity in the string fluid limit
is exactly equivalent to MHD.

\vspace{1em} \emph{MHD/string fluid correspondence.}---The dynamics of MHD is
determined by the equations of motion \eqref{EOM_MHD} where $A_\mu$ is a
dynamical gauge field. In this setting, $B^\mu$ is treated as $\mathcal{O}(1)$
while $E^\mu$ as $\mathcal{O}(\dow)$. The constitutive relations of MHD are
solutions of the offshell second law of thermodynamics
\begin{equation} \label{eq:ad}
  \nabla_\mu N^\mu_{\text{MHD}} = \half T^{\mu\nu} \delta_\scB g_{\mu\nu}
  + J^\mu \delta_\scB A_\mu + \Delta~, \quad
  \Delta \geq 0~.
\end{equation}
Recalling 
that
$J^{\mu\nu} = \half \epsilon^{\mu\nu\rho\sigma} F_{\rho\sigma}$ and
$J^\mu_\ext = \frac16 \epsilon^{\mu\nu\rho\sigma}H_{\nu\rho\sigma}$ together
with
\begin{equation}
  N^\mu = N^\mu_{\text{MHD}} - \frac{1}{T} J^{\mu\nu} \zeta_\nu
  + \frac{\mu}{T} J^\mu_\ext~,
\end{equation}
it follows that 
\eqref{eq:ad} in MHD is equivalent to
\eqref{eq_fn}, provided that Maxwell's equations, $J^\mu + J^\mu_\ext = 0$, are
taken onshell. Formally, this requires solving for $\mu$ and $E^\mu$ using
Maxwell's equations, after which the remaining fields $u^\mu$, $T$, and $B^\mu$
can be mapped to $u^\mu$, $T$, $\varpi$, and $h^\mu$ of string fluids, modulo
hydrodynamic frame transformations \cite{Armas:2018zbe}.  Given that 
identification of \eqref{eq:ad} and \eqref{eq_fn},
the equivalence between the most generic constitutive relations allowed by it
also follows.

As an example, we consider the first order charge current $J^\mu$ in
parity-invariant MHD~\cite{Hernandez:2017mch}
\begin{multline} \label{eq:current}
  J^\mu = \lb q
  {+} \frac{\dow M_\Omega}{\dow \mu}
  \epsilon^{\lambda\nu\rho\sigma} B_\lambda u_\nu \dow_\rho u_\sigma \rb u^\mu
  - \epsilon^{\mu\nu\rho\sigma} \dow_\rho (\varpi u_\nu \hat B_\sigma) \\
  + \lb \sigma_\perp \bbB^{\mu\nu}
  {+} \sigma_\parallel \hat B^\mu \hat B^\nu
  {+} \tilde\sigma \epsilon^{\mu\nu\rho\sigma} u_\rho \hat B_\sigma \rb V_\nu
  + \mathcal{O}(\dow^2)~,
\end{multline}
where $M_\Omega$ is a hydrostatic transport coefficient while $\sigma_\perp$,
$\sigma_{\parallel}$, and $\tilde\sigma$ are 
dissipative ones. The remaining
parameters obey the thermodynamic relations $dP = sd T + q d \mu -\varpi d |B|$
where $P$ is the fluid pressure in MHD. We have also defined
$\hat B^\mu = B^\mu/|B|$,
$\bbB^{\mu\nu} = g^{\mu\nu} + u^{\mu}u^{\nu} - \hat B^\mu \hat B^\nu$, and
$V^\mu = E^\mu - T P^{\mu\nu} \dow_\nu (\mu/T)$. Projecting \cref{eq:current}
along $u^\mu$, we have that $q(T,\mu, B^2)=\mathcal{O}\left(\dow\right)$, which
can be formally solved by $\mu=\mu_0(T,B^2)+\mathcal{O}\left(\dow\right)$
\cite{Note3}. In turn, one can determine $E^\mu$ by projecting \cref{eq:current}
transverse to $u^\mu$.  Comparing this result with
$F^{\mu\nu} = -\half \epsilon^{\mu\nu\rho\sigma} J_{\rho\sigma}$ together with
\cref{eq:transport}, we can identify $B^\mu = \rho h^\mu + \mathcal{O}(\dow)$
and the transport coefficients
\begin{gather}
  \alpha = \mu_0,~~
  \beta = \rho M_\Omega+\varpi\alpha,~~
  r_\perp =
  \frac{\sigma_\perp}{\sigma_\perp^2 + \tilde\sigma^2} \left(\frac{sT}{\epsilon +
      p}\right)^2,
  \nn\\
  \tilde r_\perp = \left( \frac{sT}{\epsilon + p}\right)^2 \lb \frac{-\tilde\sigma}{\sigma_\perp^2 + \tilde\sigma^2}
  + \frac{2 \rho\alpha}{sT} \rb ,~~
  r_\parallel = \frac{1}{\sigma_\parallel},
  \label{eq:map}
\end{gather}
where $p = P+ \varpi \rho$. The remaining transport coefficients in
\cref{eq:transport} can be identified by comparing the stress tensor
$T^{\mu\nu}$ in the two formulations.  Using the results of
\cite{Hernandez:2017mch}, explicitly, we find
\begin{equation}
\begin{gathered}
  \zeta_\perp = \zeta_1 - \frac23 \eta_1~, \qquad
  \zeta_\times = \zeta_1 + \zeta_2 - \frac23\eta_1 - \frac23 \eta_2~, \\
  \zeta'_\times = \zeta_1 + \frac43 \eta_1~, \qquad
  \zeta_\parallel = \zeta_1 + \zeta_2 + \frac43\eta_1 + \frac43 \eta_2~,
\end{gathered}
\end{equation}
where $\zeta_1$, $\zeta_2$, $\eta_1$, and $\eta_2$ have been defined
in~\cite{Hernandez:2017mch}. The other four coefficients $\eta_\perp$,
$\eta_\parallel$, $\tilde\eta_\perp$, and $\tilde\eta_\parallel$ map one-to-one
with those denoted by the same symbols in~\cite{Hernandez:2017mch}. Hence, we
see that at one-derivative order, MHD is entirely equivalent to one-form
superfluids in the string fluid limit.

Note that the equivalence between the two formulations, even just in the
dissipative sector, required the presence of the hydrostatic coefficient
$\alpha$ in one-form superfluids, which captures a non-zero $\mu = \mu_0$ in an
MHD configuration. This coefficient was absent in all the previous discussions
on string fluids~\cite{Schubring:2014iwa,Grozdanov:2016tdf,Hernandez:2017mch},
thereby the equivalence only holding in a state with $\mu_0 = 0$ of
MHD. Furthermore, the authors of~\cite{Hernandez:2017mch} focused only on the
dissipative sector and considered this equivalence at the linearised level
\cite{Note4}. However, here we have provided the exact
non-linear map between transport coefficients, including the hydrostatic sector
which had not been previously analysed.

\vspace{1em} \emph{Outlook.}---In this letter we have formulated a theory of
one-form superfluidity and illustrated how MHD can be understood as a particular
sector of this theory. This dual formulation is in many ways a better and
cleaner description of MHD as it makes all the global symmetries of MHD
manifest, eliminates the non-propagating fields $\mu$ and $E^\mu$, and the
electric fields being $\mathcal{O}(\dow)$ becomes a consequence rather than an
assumption. Most importantly, unlike the conventional formulation, in the
superfluid formulation, the constitutive relations are directly obtained for the
physically observable electromagnetic fields, which considerably simplifies the
computation of correlation functions.

This description of MHD as superfluidity is entirely based on conservation
equations, even when including dissipation effects. Together with the
understanding of the necessary conditions for equilibrium, it can provide
initial configurations for obtaining interesting insights in the context of
astrophysical phenomena using numerical simulations
(e.g. \cite{2010MNRAS.408..752P}). As a proof of concept, we obtain an
equilibrium configuration (i.e. without dissipation) of a slowly rotating
magnetised star on a flat background $g_{\mu\nu}=\eta_{\mu\nu}$,
$b_{\mu\nu}=0$. The equilibrium configuration corresponds to
$u^\mu/T = \delta^\mu_t + \omega (y\delta^\mu_x - x \delta^\mu_y)$ and
$\mu_\mu/T = 0$, where $\omega$ is a small angular velocity along the
$z$-axis. Assuming an equation of state $p=p_m(T)- \frac{1}{2\chi} \varpi^2$
with constant magnetic susceptibility $\chi$, we can find a solution of
\cref{vf-EOM} for the scalar Goldstone as
$\varphi_0 = - z \cos\alpha + (x - \omega ty) \sin\alpha +
\mathcal{O}(\omega^2)$, leading to
$\varpi h_\mu = \cos\alpha\, \delta_\mu^z - \sin\alpha (\delta^x_\mu - \omega y
\delta^t_\mu - \omega t \delta^y_\mu) + \mathcal{O}(\omega^2)$. The angle
$\alpha$ parametrises the misalignment of the magnetic axis (direction of
strings) with the rotational axis of the star. For a non-zero $\alpha$, the
vector $h_\mu$ is not aligned along a spacelike isometry neither with a linear combination of isometries as
in~\cite{Caldarelli:2010xz, Grozdanov:2016tdf}, and hence necessitates the new theory that we have
presented in this letter. We also note that in the traditional
formulation of MHD, one would be required to solve Maxwell's equations for
equilibrium configurations of $A_\mu$, and typically simplifications such as
working within the force-free electrodynamics (FFE) regime are imposed 
On
the other hand, using the framework proposed here, we have reduced the problem
of finding equilibrium configurations to the problem of finding solutions to a
scalar Poisson's equation \eqref{vf-EOM}. This is a considerable leap forward
towards providing appropriate initial conditions for numerical simulations in
arbitrary spacetime backgrounds. More generally, we expect this theory to be
useful for obtaining new analytic equilibrium solutions for accretion disks
surrounding Kerr black holes or magnetised stars such as pulsars, and to probe
mechanisms of energy transport therein by studying fluctuations around such
solutions including the effects of dissipation. \\

\begin{acknowledgments} 
  We would like to thank J. Bhattacharya, J. Hernandez, and specially N. Iqbal
  for various helpful discussions. We would also like to thank J. Bhattacharya,
  S. Grozdanov and N. Iqbal for comments on an earlier draft.  JA is partly
  supported by the Netherlands Organization for Scientific Research (NWO).  AJ
  would like to thank Perimeter Institute, where part of this project was done,
  for hospitality. AJ is supported by the Durham Doctoral Scholarship offered by
  Durham University.
\end{acknowledgments}

\addcontentsline{toc}{section}{References}
\footnotesize

\end{document}